\documentclass[aps,prb,twocolumn,showpacs,amsmath]{revtex4}
\usepackage[dvips]{graphicx}
\begin{document} 
\title{Two electrons in a strongly coupled double
 quantum dot: from an artificial helium atom to a hydrogen molecule}
\author{W. Dybalski$^*$}
\author{P. Hawrylak}
\affiliation{Institute for Microstructural Sciences, National Research Council
Canada, Ottawa, Canada K1A0R6, \\
and\\
 Physics Department, 
    University of Ottawa, 
    150 Louis Pasteur, 
    Ottawa, Ontario K1N 6N5 
    Canada }
\date{\today}
\begin{abstract}
We study the formation of molecular states in a two-electron quantum dot as a function of the
barrier potential dividing the dot. The increasing barrier potential  drives the two electron system 
from an artificial helium atom to an artificial hydrogen molecule. To study this strongly coupled regime, 
we introduce variational 
wavefunctions which describe accurately two electrons in a single dot, and then study their mixing 
induced by the barrier. The evolution of the singlet-triplet gap with the barrier potential and 
with an external magnetic field is analyzed. 

\end{abstract}


\newcommand{\B}{\omega_c}
\newcommand{\beq}{\begin{equation}}
\newcommand{\eeq}{\end{equation}}
\newcommand{\beqa}{\begin{eqnarray}}
\newcommand{\eeqa}{\end{eqnarray}}
\newcommand{\om}{\omega}
\newcommand{\Om}{\Omega}
\newcommand{\fr}{\frac}
\newcommand{\lan}{\langle}
\newcommand{\ran}{\rangle}
\newcommand{\De}{\Delta}
\newcommand{\eps}{\epsilon}
\newcommand{\rj}{\vec{r}_1}
\newcommand{\rd}{\vec{r}_2}
\newcommand{\Rv}{\vec{R}}
\newcommand{\rv}{\vec{r}}
\newcommand{\omk}{\omega_0^2}
\newcommand{\pa}{\partial}
\newcommand{\vp}{\varphi}
\newcommand{\al}{\alpha}
\newcommand{\sal}{\sqrt\al}
\newcommand{\be}{\beta}
\newcommand{\spi}{\sqrt{\pi}}
\newcommand{\spa}{\sqrt{\fr{\pi}{\al}}}
\newcommand{\ali}{\fr{1}{\sqrt{\al}}}
\newcommand{\Ht}{\tilde{H}}
\newcommand{\Ga}{\Gamma}
\newcommand{\ga}{\gamma}
\newcommand{\de}{\delta}
\newcommand{\tbe}{\tilde{\be}}
\newcommand{\tde}{\tilde{\de}}
\newcommand{\tu}{\tilde{u}_{0,\tbe}}
\newcommand{\tC}{\tilde{C}_{0,\tbe}}
\newcommand{\Dj}{\Delta_1}
\newcommand{\Dd}{\Delta_2}
\newcommand{\tuo}{\tilde{u}_{0,\tbe_0}}
\newcommand{\tCo}{\tilde{C}_{0,\tbe_0}}
\newcommand{\tE}{\tilde{E}}

\maketitle

\section{Introduction}
There is currently interest in developing means of isolating spins of individual 
electrons and coupling them in a controlled 
way \cite{jose-pawel,loss-divincenzo,burkard-loss,hu-dassarma1,WSSR,michel-ddotprl,kouwenhoven,leburton-lddot,Michel}. 
This problem is equivalent to a formation and controlled dissociation of an
artificial hydrogen molecule. This dissociation is not achieved by the increase of the
separation of the hydrogen atoms, but by the increase of the tunneling 
barrier separating the dots. In such a process a single two-electron dot breaks 
into two one-electron dots.
Hence the analogy to the break-up of the helium atom into the hydrogen molecule,
or to nuclear fission, rather than to a chemical reaction.
The description of the "artificial fission" process cannot be accomplished by
the weak coupling approaches, such as the Hund-Mulliken method, and requires
the treatment of a strongly coupled electron system. We develop such an approach
in this work.

The quantum mechanical study of two electron atoms dates back
to the work of Born and Heisenberg \cite{BH} on the helium atom.
(See \cite{Bethe} for a review on two-electron atoms).
Similar studies of artificial atoms  followed
the development of quantum dots \cite{JHW,RCA,RM} - 
nanostructures in which the number of electrons can be reduced to a desired value 
($n_e=0,1,2...$) in a controllable manner. In contrast to atoms, 
the confining potential of a quantum dot is, to a good approximation, 
quadratic, so a dot containing one electron provides a 
realization of the exactly soluble Fock-Darwin model \cite{Fock,Darwin,JHW,Raymond}. 

The problem of two interacting electrons in a parabolic potential also admits exact
solutions, but only for specific values of the oscillator frequency, as was shown by
Taut \cite{Taut,Taut1,Taut2}.
The general case was treated analytically by the oscillator representation method \cite{Dineykhan}
and variational calculations \cite{Das}, and  studied numerically by the following  approaches:
'exact' diagonalization using Fock-Darwin states \cite{PGM,JMM}, integration
 of the radial motion Schroedinger equation after separating the center of 
 mass motion \cite{WMC} and a combination of both \cite{PH}. 
The results were compared with experimental data \cite{JMM,PH}
 and with the Hartree and Hartree-Fock methods \cite{PGM}. 
The mean field approaches were
 applied to the two-electron system also in the context of symmetry breaking they may 
 induce  and subsequent symmetry restoration by RPA \cite{SNP} and projection techniques
 \cite{Uzi}. This theoretical problem, as well as the exact and numerical solutions
 mentioned above, proved relevant for a description of a Wigner molecule consisting of 
 two electrons in a quantum dot \cite{PSN,Uzi1}. 

The problem of two vertically \cite{Palacios,manfred-science,austing_molinari,peeters,
leburton-vddot} 
or laterally coupled 
dots \cite{jose-pawel,loss-divincenzo,burkard-loss,hu-dassarma1, WSSR, leburton-lddot} 
containing one electron each is equivalent to the problem of an 
artificial hydrogen molecule. 
A variety of methods were applied here: The general case was studied by 
LSDA \cite{WSSR}, molecular orbital calculations \cite{burkard-loss,hu-dassarma1} and  
the Hartree-Fock approach \cite{hu-dassarma1,Uzi,Uzi2,Uzi3} refined subsequently by 'exact' 
diagonalization \cite{hu-dassarma2} and projection techniques \cite{Uzi,Uzi2,Uzi3}.
The weakly coupled regime  was studied analytically by the Heitler-London and 
Hund-Mulliken methods \cite{loss-divincenzo,burkard-loss}. 
While the analytical results by the molecular Heitler-London and 
Hund-Mulliken approaches \cite{burkard-loss} are very useful,
the weakly coupled regime does not quantitatively describe the experimental situation.
In molecular description the starting point are two well separated quantum dots. 
Then, as the distance between them is reduced, the electrons start tunneling from one 
dot to another in analogy to a chemical bond formation. 
However, in an actual experiment the double lateral dot is defined electrostatically 
by metallic gates located above the two dimensional electron 
gas \cite{michel-ddotprl,kouwenhoven}. Here the distance between the dots 
is held fixed and the coupling between them is controlled by means of the inter-dot       
barrier. When the barrier is zero the electrons move freely  and
our system is a single dot, an artificial helium atom. When the barrier increases,  the
single dot divides into two, the electrons reconfigure so as to avoid the
barrier and an artificial hydrogen molecule forms. 
It is difficult to find an analogue of the above procedure
in the realm of atomic physics. Nuclear physics, however, offers an obvious 
example - fission of a nucleus. In this work we demonstrate that viewing quantum
dots as 'artificial nuclei' rather than 'artificial atoms' offers also some
computational advantages.

Our paper is organized as follows: Section II describes our model consisting
of a two-dimensional parabolic potential perturbed by a Gaussian 
barrier running along its diameter. In Subsection IIA we briefly discuss 
the exact eigenvectors of the single dot problem, found by Taut \cite{Taut2},
which are, however, correct only for specific values of the magnetic field, 
different for each state. In Subsection IIB  
we introduce variational wavefunctions which reduce to the exact 
eigenvectors at these specific magnetic fields. We calculate
corresponding variational energies and compare them with exact and numerical values.
In Section III we switch on the barrier and describe the 
formation of molecular states localized in the two potential minima, and the effect of
the magnetic field on the singlet-triplet gap.

\section{The model Hamiltonian}
Our model Hamiltonian describes two electrons moving in the $(x,y)$ plane, confined
by a parabolic potential with frequency $\om_0$,  perturbed by 
a Gaussian barrier of width $\De$ and hight $V_0$, and subject to a perpendicular
magnetic field $\vec{B}$:
\begin{widetext}
\beq
H=-(\vec{\nabla}_1+i\vec{A}(\rj))^2-(\vec{\nabla}_2+i\vec{A}(\rd))^2
+\fr{1}{4}\omk(\rj^2+\rd^2)+\fr{2}{|\rj-\rd|}+
V_0(e^{-\fr{x_1^2}{\De^2}}+e^{-\fr{x_2^2}{\De^2}}). \label{full}
\eeq
\end{widetext}
Here $\vec{A}(\vec{r}_i)=(\B y_i/4,-\B x_i/4,0)$, $\B=\fr{eB}{m^*}$ is the cyclotron
frequency of an electron with effective mass $m^*$ and charge $-e$ placed in an external
magnetic field $B$. 
The lengths are expressed in effective Bohr radii $a_B=\fr{4\pi\epsilon\hbar^2}{m^* e^2}$
(where $\epsilon$ is the electric permeability), whereas $\om_0$, $\B$ and $V_0$ 
in effective rydbergs ($1Ry=\fr{\hbar^2}{2 m^* a_B^2}$). The magnetic field 
points in the negative direction of the $z$ axis.


\subsection{Special exact solutions of the single dot problem} 

In this and the following subsection we set $V_0=0$ and address the single dot problem. 
Change of variables $\Rv=\fr{\rj+\rd}{2}$, $\rv=\rd-\rj$ separates the 
relative and center of mass motion in the Hamiltonian: $H=H_R+H_r$, where $H_R$ is the 
Fock-Darwin one-particle Hamiltonian. 
The detailed study of the radial Hamiltonian $H_r$ is presented in \cite{Taut2}. 
Here we only summarize the results which we will need in the sequel: The relative motion 
eigenfunction with angular momentum $m$ can be expressed as follows:
\beq
\psi(\rv)=\sqrt[4]\al\fr{u_m(\rho)}{\sqrt{\rho}}\fr{e^{im\phi}}{\sqrt{2\pi}},
\eeq
where $\al=\fr{1}{4}\sqrt{\omk+\fr{\B^2}{4}}$, $\rho=\sqrt{\al}r$. 
Aiming at the lowest excitations of the radial motion one obtains
\beq
u_{m}(\rho)=C_m\rho^{\fr{1}{2}+|m|}(1+\sqrt{\fr{2}{1+2|m|}}\rho)e^{-\fr{\rho^2}{2}},
\label{wf}
\eeq
(where $C_m$ are the normalization constants) and the corresponding eigenenergy:
\beq
E_m=-\fr{\B m}{2}+2\fr{(2+|m|)}{1+2|m|}.                                               
\eeq                                                                                        
However, the two expressions above are valid only if
\beq
\al=\fr{1}{2(1+2|m|)} \label{al}.
\eeq
In particular, for $m=0$ we get $E_0=4 Ry$ provided that $\al=\fr{1}{2}$, and consequently
$\B=\omega_{c 0}=2\sqrt{4-\omk}$. The corresponding wavefunction reads:
\beq
u_0(\rho)=C_0\rho^{\fr{1}{2}}(1+\sqrt 2\rho)e^{-\fr{\rho^2}{2}}\label{0}.
\eeq
At zero magnetic field this is the ground state wavefunction, as it is nodeless. 
For $m=\pm 1$ the energy equals $E_{\pm 1}=\mp\fr{\B m}{2}+2\: Ry$, 
on condition that $\al=\fr{1}{6}$. This
implies that $\B=\omega_{c 1}=2\sqrt{\fr{4}{9}-\omk}$. 
The triplet radial wavefunction follows:
\beq
u_1(\rho)=C_1\rho^{\fr{3}{2}}(1+\sqrt{\fr{2}{3}}\rho)e^{-\fr{\rho^2}{2}}\label{1}.
\eeq
For example, if $\omega_0=\fr{2}{3} Ry$ then the triplets have the energy 
2Ry at zero magnetic field, whereas the lowest singlet has the energy 4Ry
at $\B=\fr{8\sqrt{2}}{3}\approx 3.77Ry$. But we are not able to get the
exact energies at intermediate values of $\B$. The subject of the next
section will be a derivation of accurate upper bounds for these energies.

\subsection{Variational analysis}

\begin{figure}
\begin{center}
\includegraphics[angle=0,width=0.5\textwidth]{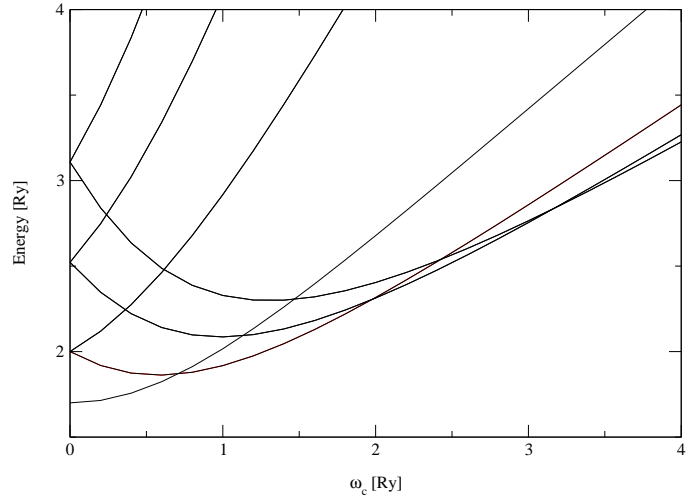}  
\caption{Comparison of the relative motion energies in a parabolic
dot with $\omega_0=\fr{2}{3} Ry$ calculated variationally and by the Numerov method. 
Solid lines: variational m=0,1,-1,2,-2,3,-3 states 
(increasing order at $\B=0.1 Ry$).
Dashed line: numerical m=0 state. Other numerical states coincide
with variational in this figure.}
\label{sdot}
\end{center}
\end{figure}

To describe the lowest lying states of radial motion with arbitrary angular 
momentum we suggest  variational wavefunctions inspired by the form 
of the eigenfunction (\ref{wf}). It does not seem reasonable to change
the factors $\rho^{\fr{1}{2}+|m|}$ or $e^{-\fr{\rho^2}{2}}$ as it would 
spoil the behaviour of the function  at zero or at infinity.
The only remaining parameter is the one that multiplies $\rho$ in the 
bracket. Therefore, we introduce the following family of variational
wavefunctions, labelled by the parameter $\be$:     
\beq
u_{m,\be}(\rho)=C_{m,\be}\rho^{\fr{1}{2}+|m|}(1+\be\rho)e^{-\fr{\rho^2}{2}}.                   
\eeq
The corresponding variational energies follow:
\beq
E_m(\be)=-\fr{\B m}{2}+2\al\fr{a_m+b_m\be+c_m\be^2}{d_m+e_m\be+f_m\be^2}, \label{ve2}
\eeq
where $d_m=\fr{1}{2}\Ga(1+|m|)$, $e_m=\Ga(\fr{3}{2}+|m|)$, $f_m=\fr{1}{2}\Ga(2+|m|)$, 
\beqa
a_m &=& \fr{e_m}{(2|m|+1)\sqrt{\al}}+2f_m, \nonumber\\
b_m &=& \fr{2d_m}{\sqrt{\al}}+2(|m|+1)e_m,  \nonumber\\
c_m &=& \fr{e_m}{2\sqrt{\al}}+(2m^2+4|m|+3)d_m. \nonumber
\eeqa
The minimum $\be_m$ of this simple function is found to be at one of the roots of the
quadratic equation:
\beqa
(b_m d_m-a_m e_m)+2(c_m d_m-a_m f_m)\be+ \nonumber \\
(c_m e_m -b_m f_m)\be^2=0. \label{quadraticequation}
\eeqa
There arises a question of accuracy of our method.
To study this problem let us return to the example presented in the end of
the previous subsection where $\omega_0=\fr{2}{3}\:Ry\approx 4meV$ in GaAs. 
Clearly, our variational energies reproduce the exact eigenenergies 
at the specific values of $\B$. (In this respect they improve upon the interpolation 
formula of Taut \cite{Taut2}.) To evaluate accuracy at intermediate values of $\B$ 
we solve the radial eigenvalue problem numerically, by the Numerov method, and compare the results 
with our variational energies (Fig. \ref{sdot}).
The picture shows that variational and numerical 
calculations are in very good agreement in a wide range of the magnetic field. 
The $m\ne 0$ energies calculated by both methods give almost identical results 
and are represented by single solid lines. The variational $m=0$ energy is even 
more accurate than the
numerical result, as it is below the numerical value for large magnetic
fields. Lower accuracy of the Numerov method is probably due to the fact that 
m=0 state is the only one which is non-zero in the singular point $\rho=0$.

In Fig. \ref{sdot} we note the familiar singlet-triplet oscillations
of the ground state \cite{WMC,JMM,RCA,Uzi3,Scarola}. They are caused by a combination 
of two mechanisms:
First, because of the orbital Zeeman term $-\fr{\B L_z}{2} $ it is energetically favourable for
the system to rotate. Nonetheless, it is clear that without interaction the
ground state would be a singlet in any magnetic field. Second, Coulomb interaction
acts stronger on singlets than on triplets, as in the latter the electrons are kept
apart by the Pauli principle. Thereby, the gap between a singlet and a consecutive triplet
is reduced below its non-interacting value. Nonetheless, without the Zeeman term
the ground state is always a singlet no matter how strong is the interaction 
or the term quadratic in $\B$ \cite{RS}.
We conclude that, in the presence of interaction, the increase of the magnetic field results 
in an increase of ground state angular momentum. As states with even $m$ are singlets, 
whereas states with odd $m$ are triplets, it causes singlet-triplet oscillations.

\section{Double dot - formation of molecular states} 

In this section we study the ground state and the first excited state energies 
and densities of two
electrons in a double dot and their evolution with a magnetic field. 
For this purpose we go back to our model Hamiltonian (\ref{full}) and set
$V_0>0$. Clearly, the potential of the barrier: 
\beq
V_b=V_0(e^{-\fr{x_1^2}{\De^2}}+e^{-\fr{x_2^2}{\De^2}})
\eeq
couples the motion of the center of mass and the relative motion.
Nevertheless, we choose variational wavefunctions as products:
\beqa
U_{m}(\rho,\vp,\vec{R})=u_m(\rho,\vp)\Psi_0(\vec{R}),\\ 
u_m(\rho,\vp)=u_{m,\be_m}(\rho)\fr{e^{im\vp}}{\sqrt{2\pi}}, 
\eeqa
where the center of mass wavefunction is just the Fock-Darwin ground state: 
\beq
\Psi_0(\vec{R})=2\sqrt{\fr{\al}{\pi}}e^{-2\al R^2}. 
\eeq
The corresponding energy of the center of mass motion is
$E_{cm}=\sqrt{\om_0^2+\fr{\B^2}{4}}$.
This choice is justified by the fact that the barrier couples only every second 
center of mass wavefunction as a result of parity conservation. 
 
\begin{figure}
\begin{center}
\includegraphics[angle=0,width=0.4\textwidth]{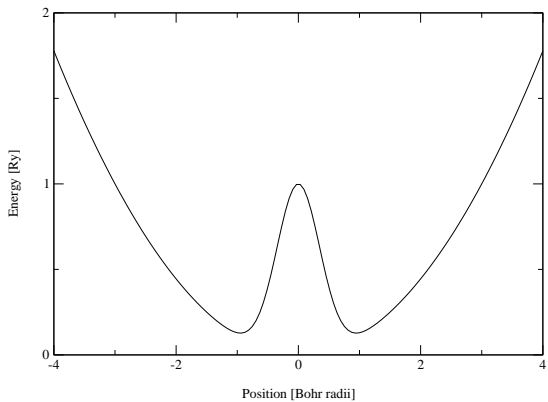}  
\caption{The profile of the double dot potential with $\omega_0=\fr{2}{3}\:Ry$, $V_0=1\:Ry$, 
$\Delta=0.5\:a_B$.}
\label{profile}
\end{center}
\end{figure}

 It is our  goal to find matrix elements
$H_{m,n}:=\lan U_m|H|U_n \ran$. To this end, we  calculate the effective potential
$V_{eff}(\rho,\vp)=\lan \Psi_0|V_b|\Psi_0\ran$ which acts only on the relative motion 
coordinates:
\beq
V_{eff}(\rho,\vp)=\fr{2V_0\ga}{\sqrt{\ga^2+1}}e^{-\fr{\rho^2cos^2\vp}{\ga^2+1}},
\label{effectivepotential}
\eeq
where $\ga=2\De\sqrt{\al}$. Now we evaluate the matrix elements of the effective
potential:
\begin{widetext}
\beqa
\lan u_m|V_{eff}|u_n\ran=V_0 C_{m,\be_m}C_{n,\be_n}\fr{\ga}{\sqrt{\ga^2+1}}
\big(K(\ga^2+1,\fr{2+|m|+|n|}{2},n-m)+\nonumber \\ (\be_m+\be_n)K(\ga^2+1,\fr{3+|m|+|n|}{2},n-m)+
\be_m\be_n K(\ga^2+1,\fr{4+|m|+|n|}{2},n-m)\big), 
\eeqa
\end{widetext}
where $K(x,k,l)=\Ga(k)\fr{1}{2\pi}\int_0^{2\pi} \fr{e^{il\vp}}{(1+\fr{1}{x}cos^2{\vp})^k}d\vp$.
For some values of $k,l$ this integral can be expressed in terms of elementary and elliptic
functions. In practice, however, we evaluate it numerically. Finally, we have the desired expression
for matrix elements of $H$:   
\beq
H_{m,n}=(E_{m}(\be_m)+E_{cm})\de_{m,n}+\lan u_m|V_{eff}|u_n\ran. \label{ddoth}
\eeq
We focus attention on the potential with parameters $\om_0=\fr{2}{3}\:Ry$, $V_0=1Ry$, $\De=0.5\:a_B$.
(The profile is shown in Fig. \ref{profile}).

In the two subsections that follow we study the behaviour of the lowest lying 
singlet and triplet energies and densities at $0 \leq \B \leq 1$. The third
subsection describes the effect of the barrier on the singlet-triplet gap.  

\subsection{Energies and wavefunctions of singlets}
\begin{figure}
\begin{center}
\includegraphics[angle=0,width=0.4\textwidth]{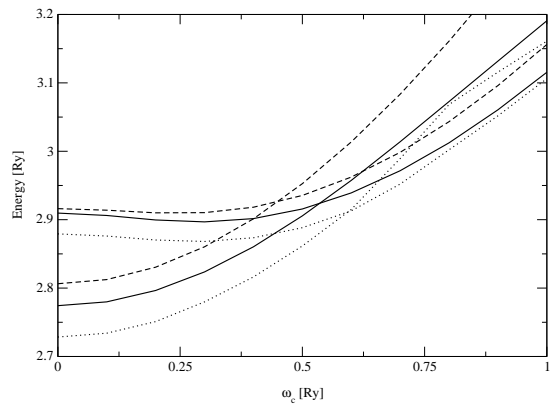}  
\caption{The lowest lying singlet and triplet energies in a double dot 
with $\omega_0=\fr{2}{3}\:Ry$, $V_0=1\:Ry$, $\Delta=0.5\:a_B$.
Dashed lines: $\tE_0$ - singlet, $\tE_{\pm 1}$ - triplet. 
Solid lines: $\tE_{0,\pm 2}$ - singlet, $\tE_{\pm 1,\pm 3}$ - triplet.
Dotted lines: numerical singlet and triplet energies calculated by the
configuration-interaction method. (In increasing order of energy
at $\B=0$).} 
\label{ddot}
\end{center}
\end{figure}

As a first approximation of the singlet energy
we take only the $u_0$ state under consideration. The resulting energy:
\beq
\tE_0=E_0(\be_0)+E_{cm}+\lan u_0|V_{eff}|u_0\ran
\eeq
is plotted as a function of the magnetic field in Fig. \ref{ddot} and compared 
with numerical calculations performed using the configuration-interaction method.  

Our variational energy is about 0.08 Ry higher than the numerical one at $\B=0$
and the discrepancy increases to 0.15 Ry at $\B=1Ry$. The reason for that 
is clear from Fig. \ref{sdot}: about $\B=1 Ry$ the $m=0$ state is close to 
degeneracy with $m=2$, therefore the latter should also be taken into account. 
On the other hand, 
$m=2$ state is degenerate with $m=-2$ at $\B=0$ so, in fact, we should consider
both of them.  We diagonalize the
resulting 3x3 matrix and note that its lowest eigenvalue $\tE_{0,\pm2}$ differs from the 
numerical result by about 0.05 Ry  at zero magnetic field (see Fig. \ref{ddot})
and this discrepancy becomes even smaller at larger $\B$. 
We recall that $m\neq 0$ states vanish when the distance between the electrons is zero.
Consequently, this increase in accuracy is 
the first manifestation of the formation of molecular states: contribution
of $m=\pm 2$ states pushes the electrons apart and, by a non-trivial angular dependence,
locates them in the dots. To support this statement we calculate the
two-electron density $\varrho $ in this approximation. Let us
denote by $\Phi(\rv_1,\rv_2)$ the two-electron wavefunction. Then:
 \beq
\varrho(\rv_1)=2\int|\Phi(\rv_1,\rv_2)|^2 d\rv_2.  \label{density}
\eeq

\begin{figure}
\begin{center}
\includegraphics[angle=-90,width=0.41\textwidth]{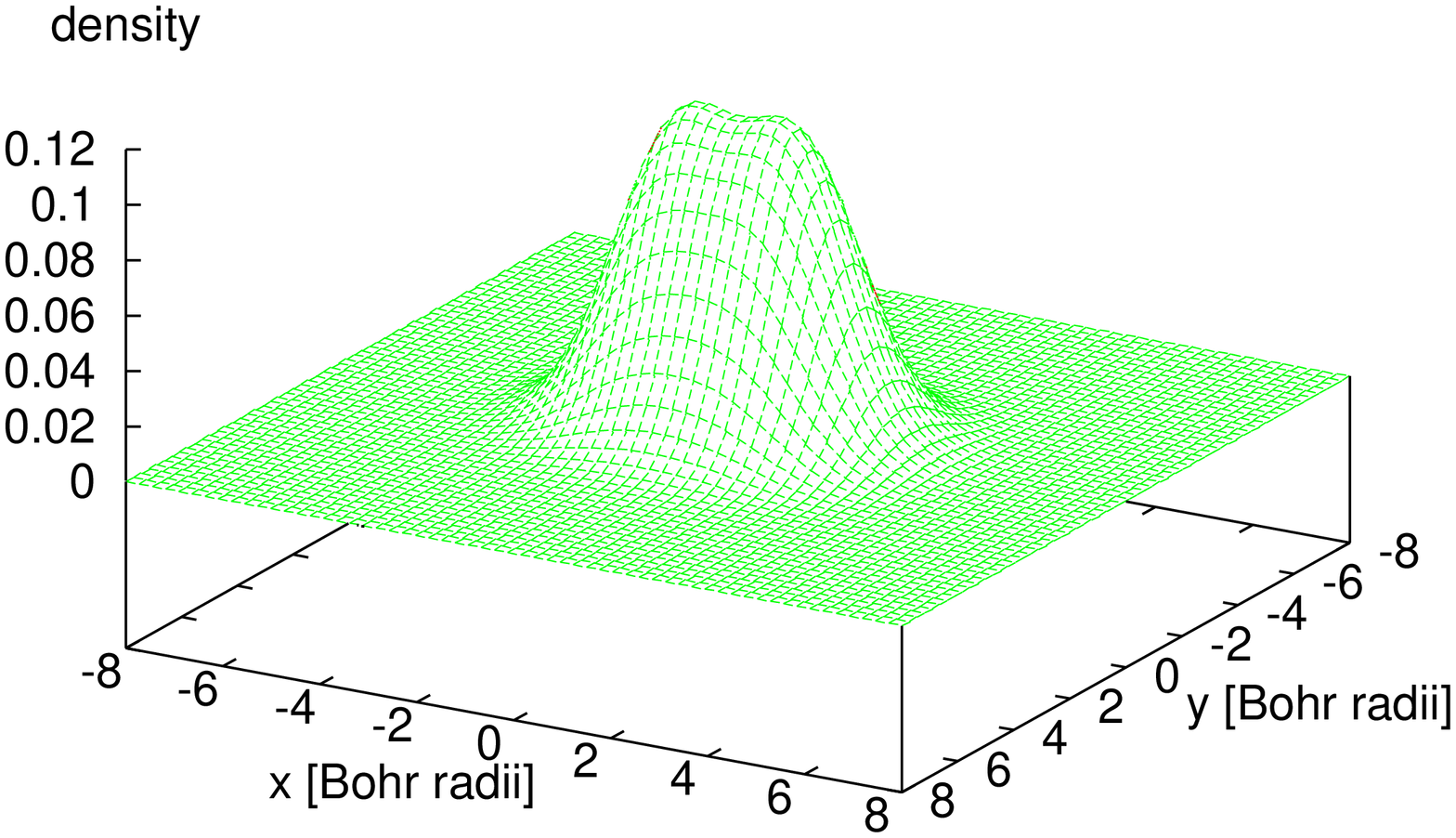} 
\includegraphics[angle=-90,width=0.41\textwidth]{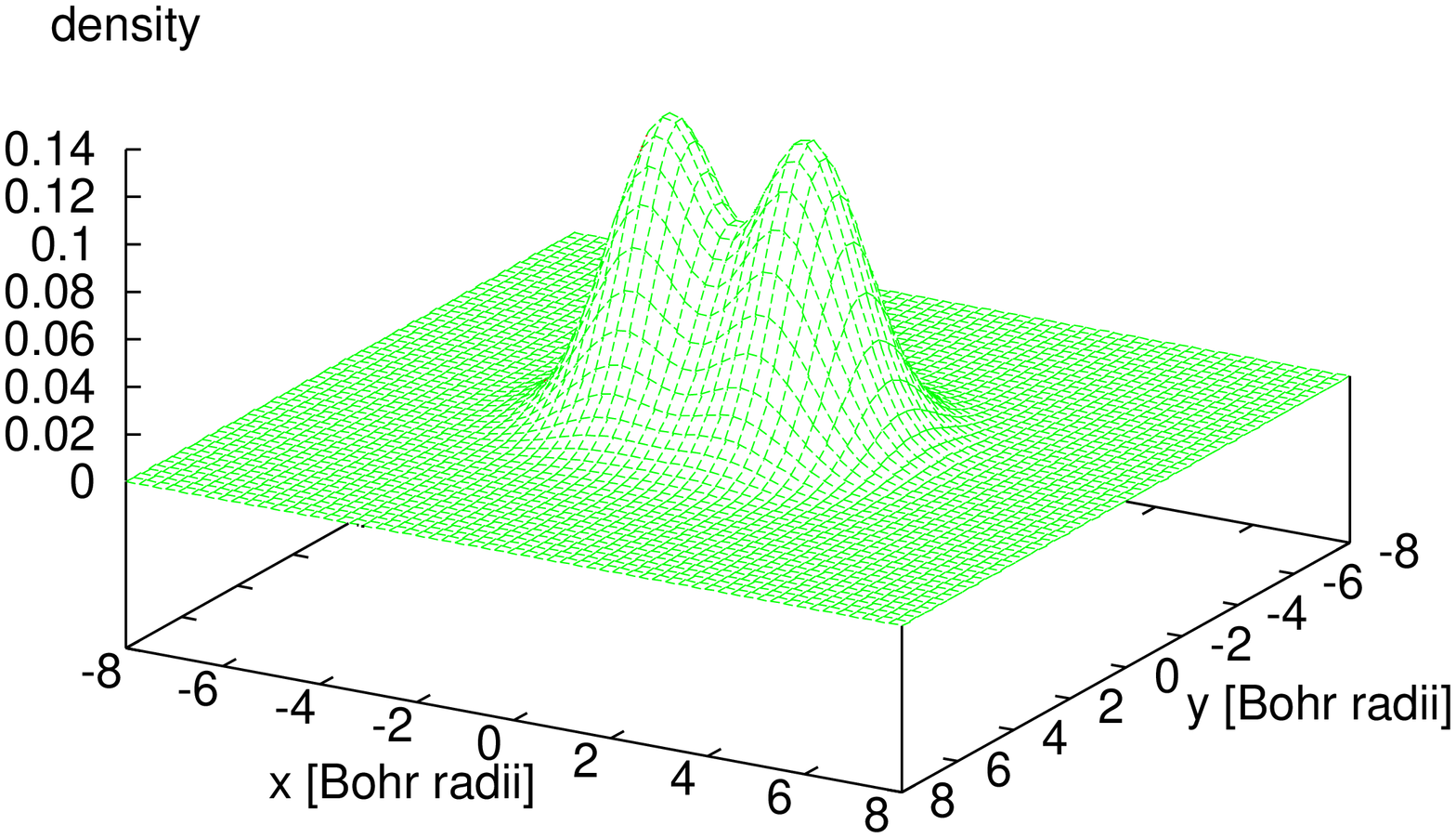}
\caption{The two-electron density of the lowest singlet calculated by diagonalization  
using $m=0,\pm 2$ states at $\B=0\:Ry$ (upper fig.), $1\:Ry$ (lower fig.).}
\label{densitiesv}
\includegraphics[angle=-90,width=0.41\textwidth]{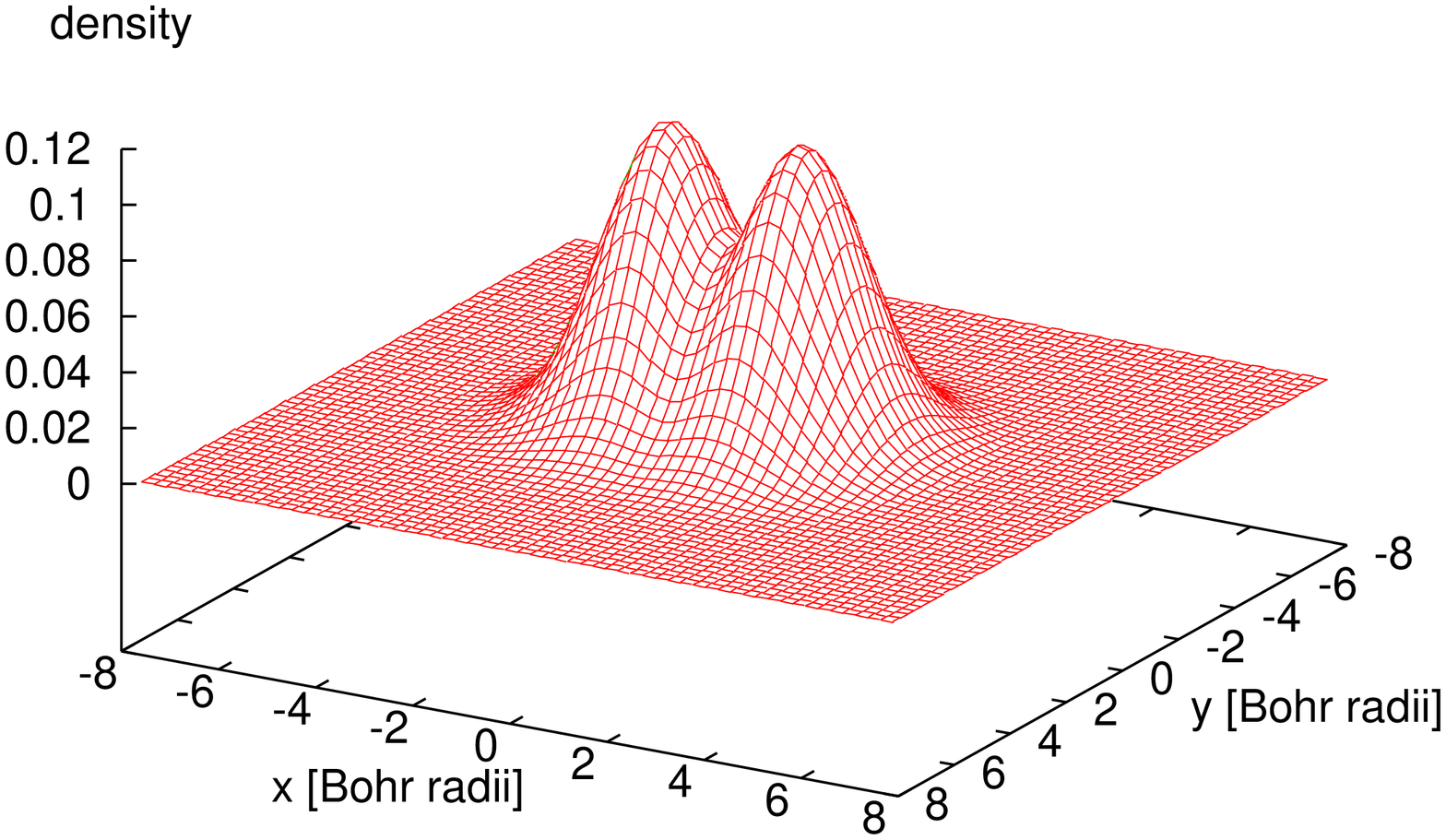} 
\includegraphics[angle=-90,width=0.41\textwidth]{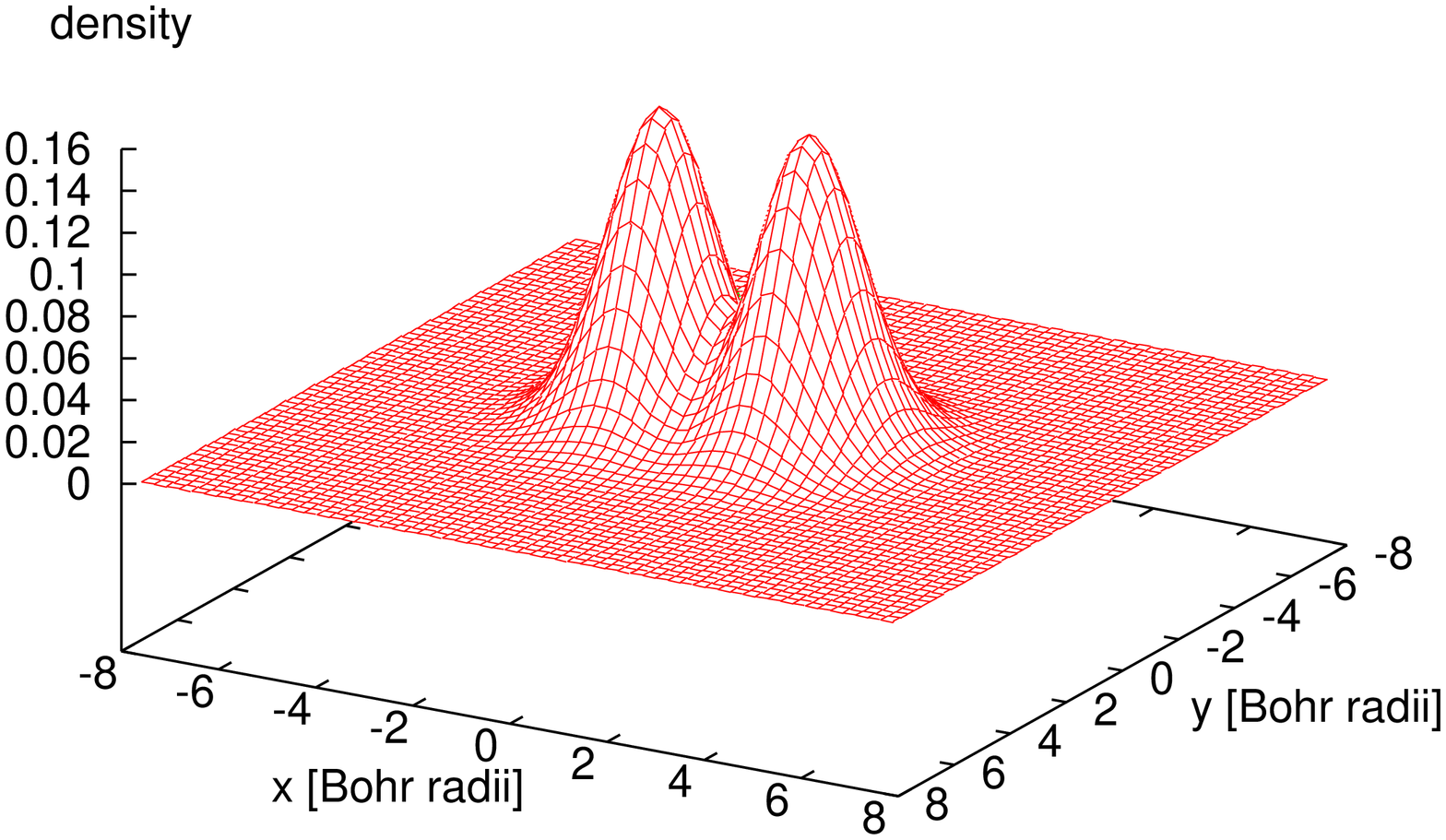}
\caption{The two-electron density of the lowest singlet calculated by the configuration-interaction method,
at $\B=0\:Ry$ (upper fig.), $1\:Ry$ (lower fig.).}
\label{densitiesn}
\end{center}
\end{figure} 
\begin{figure}
\begin{center}
\includegraphics[angle=-90,width=0.41\textwidth]{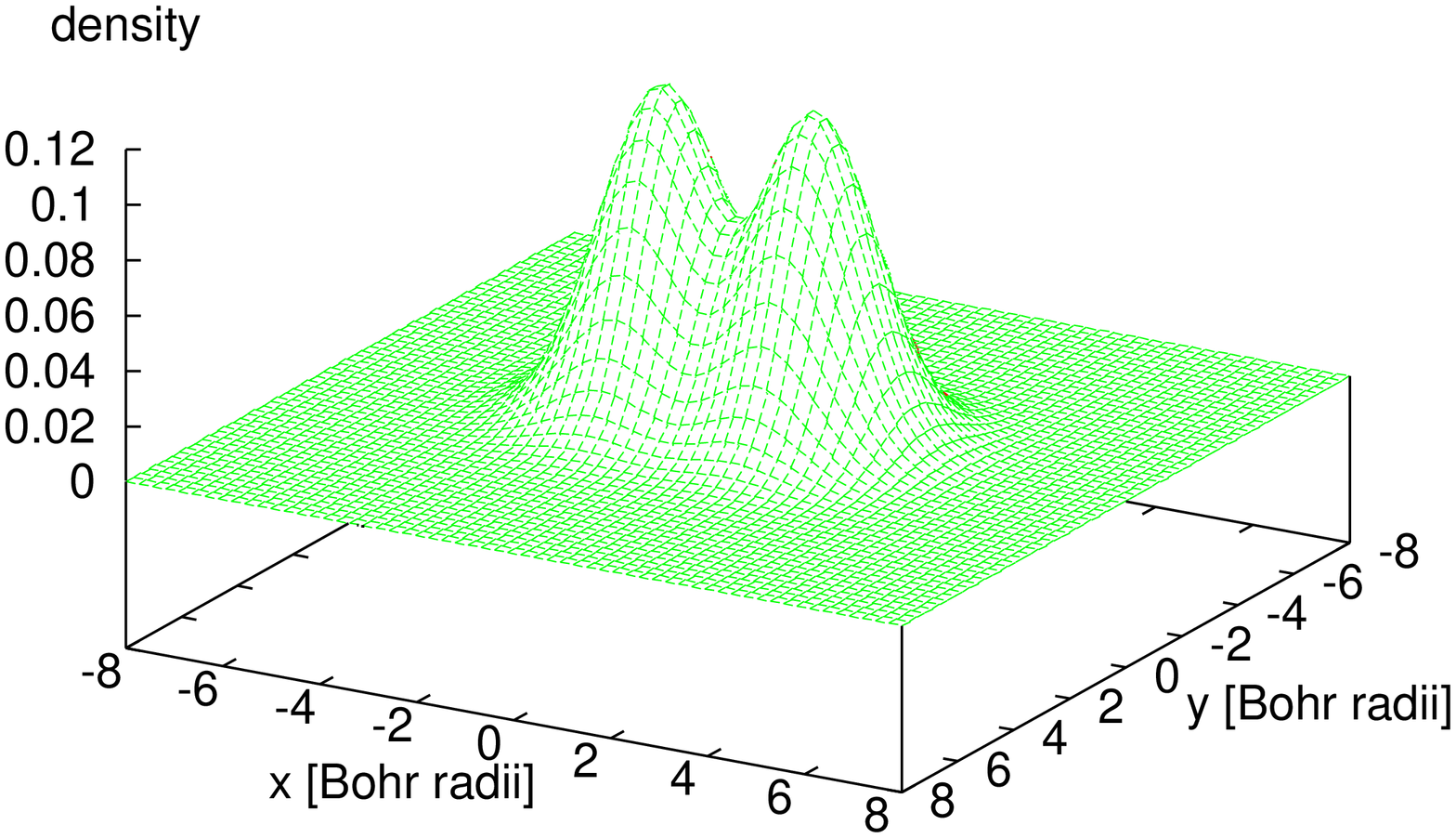}
\includegraphics[angle=-90,width=0.41\textwidth]{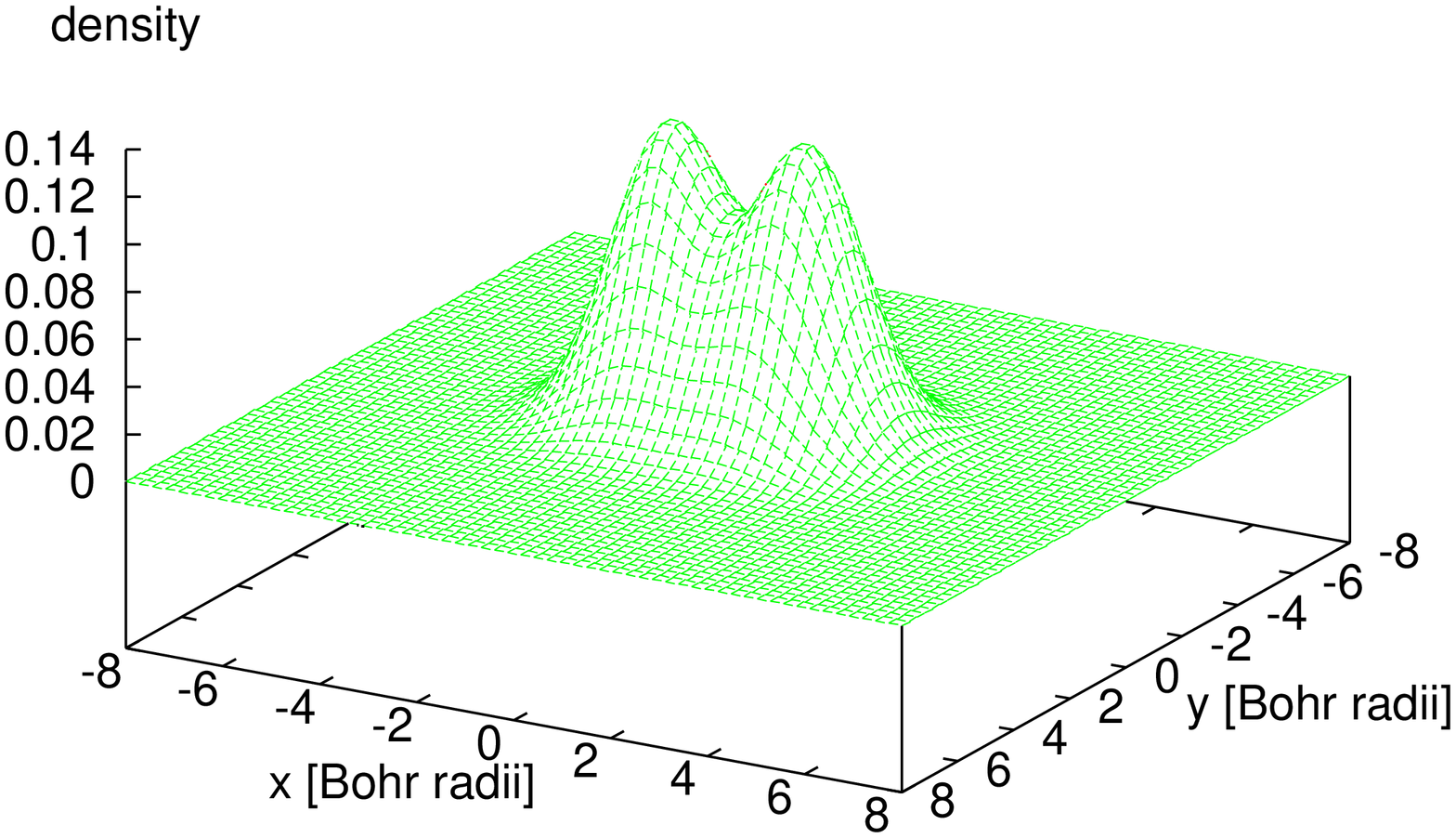}
\caption{The two-electron density of the lowest triplet calculated by diagonalization  
using $m=\pm 1,\pm 3$ states at $\B=0\:Ry$ (upper fig.), $1\:Ry$ (lower fig.).}
\label{densitiesvt}
\includegraphics[angle=-90,width=0.41\textwidth]{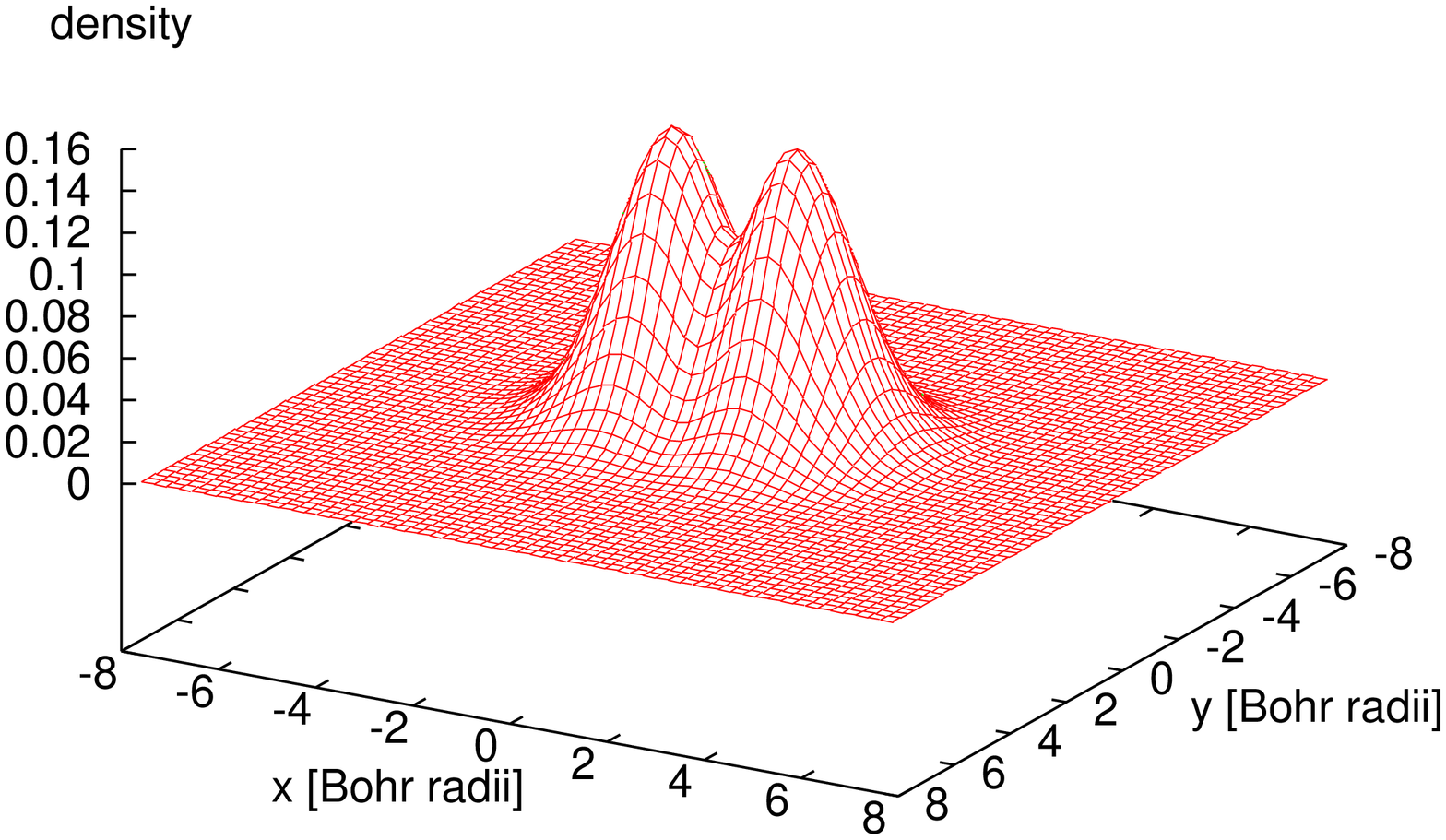} 
\includegraphics[angle=-90,width=0.41\textwidth]{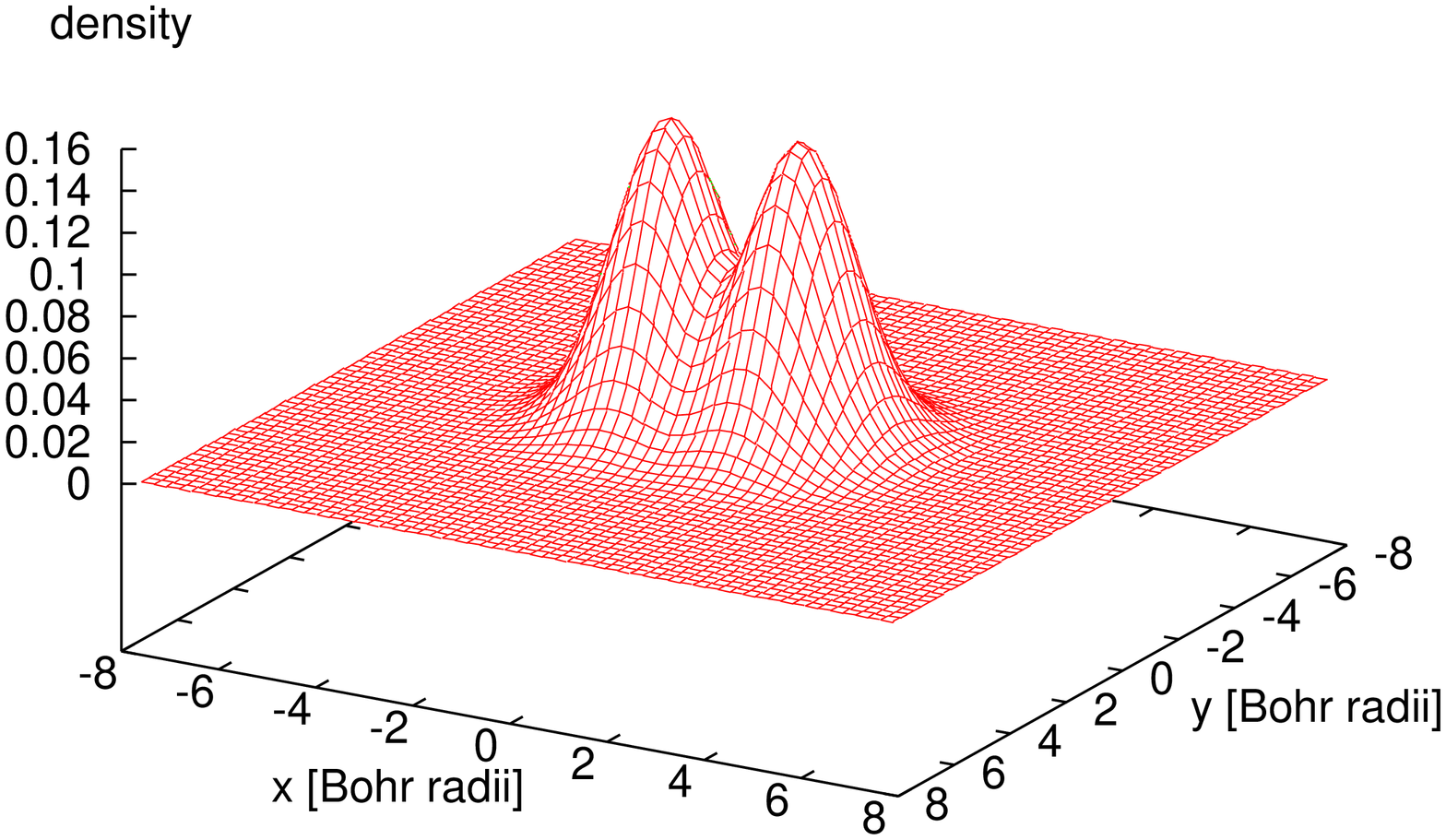}
\caption{The two-electron density of the lowest triplet calculated by 
the configuration-interaction method at 
$\B=0\:Ry$ (upper fig.), $1\:Ry$ (lower fig.).}
\label{densitiesnt}
\end{center}
\end{figure}

We have calculated the wavefunction as a linear combination of functions $U_m(\rho,\vp)$,
with the coefficients $A_m$ found by numerical diagonalization. Recalling that
 $\rho=\sqrt{\al}r$, $\rv=(x,y)=\rv_2-\rv_1$ we obtain:
\beqa
\Phi(\rv_1,\rv_2)=\sum_m A_m C_{m,\be_m}\fr{\sqrt{2}}{\pi} \al^{\fr{|m|+2}{2}}(x+iy)^m
\cdot\nonumber\\ \cdot(1+\be \sqrt{\al} r)e^{-\al(\rv_1^2+\rv_2^2)}.
\eeqa
The integral (\ref{density}) is evaluated numerically at $\B=0\:Ry$ and 
$1\: Ry$. The results are compared with densities obtained by the 
configuration-interaction method. 
(See Fig. \ref{densitiesv}, Fig. \ref{densitiesn}). We note that the
configuration-interaction calculations give a significantly lower density 
in the center of the dot than our variational method, especially at $\B=0$. This can
partly be attributed to the fact that we neglected the radial motion 
and center of mass excitations. The first excited state of radial motion
with $m=0$ is close in energy to the $m=\pm 2 $ states (see
Fig. \ref{excitedenergy}) and, in fact, lowers the density in the center of the dot 
as we show in the Appendix (see Fig. \ref{exciteddensity}). On the other hand, the center of mass
excitations which couple to the lowest singlet are separated by about $0.5\:Ry$
from $m=\pm 2$ states and we will not discuss them further.

Both methods predict that the electron density in the center of the double dot
decreases with the magnetic field. Similarly, in the Heitler-London approach \cite{burkard-loss} 
one observes that the overlap between the left and right dot wavefunction decreases 
as a function of $\B$. From our perspective the explanation of this effect starts 
from the situation in a single dot. There, as the magnetic field increases, the lowest singlet 
has larger and larger angular momentum. Consequently, the density has a circular 
ring-like shape, with a minimum in the center, caused by the centrifugal barrier. 
Now, when the inter-dot barrier is switched on, the ring shrinks into two peaks 
as a result of mixing of angular momenta and radial excitations. 

Summarizing our discussion in physical terms, the two-electron droplet in the lowest 
singlet state acquires rotating and vibrating components, when the barrier is 
increased and a magnetic field applied. The vibrations have to be included at low magnetic 
fields to obtain qualitatively correct two-electron density. They counterbalance the
non-rotating (m=0), peaked in the barrier, component of the droplet.    

\subsection{Energies and wavefunctions of triplets}
Let us now describe the lowest lying triplet. First, we remark that the 
triplet energy, as a function of the magnetic field, will have a vanishing slope
at $\B=0$. In fact, breaking of circular symmetry will eliminate the 
degeneracy of $m=\pm 1$ states at $\B=0$, making the wavefunction $\Phi$ of the lowest 
triplet real (up to a constant complex phase). Linear dependence of  triplet energy could
originate only from the orbital Zeeman term. 
But $\lan\Phi|L_z\otimes I+I\otimes L_z|\Phi\ran=0$ for
any real $\Phi$. Now we illustrate this general argument with a calculation of 
actual triplet energies taking only $m=\pm 1$ states under consideration. In this
approximation the matrix elements of the Hamiltonian (\ref{ddoth}) read:
\beqa
H_{1,1} &=& E_{1}(\be_1)+E_{cm}+\lan u_1|V_{eff}|u_1\ran, \label{diagone}\\
H_{-1,-1} &=& E_{-1}(\be_1)+E_{cm}+\lan u_{-1}|V_{eff}|u_{-1}\ran, \label{diagtwo}\\
H_{1,-1} &=& \lan u_1|V_{eff}|u_{-1}\ran.\label{nondiag}
\eeqa
We recall from Subsection IIB that $E_{1}(\be)=-\fr{\B}{2}+F(\be)$ and 
$E_{-1}(\be)=\fr{\B}{2}+F(\be)$, where 
$F(\be)=2\al\fr{a_1+b_1\be+c_1\be^2}{d_1+e_1\be+f_1\be^2}.$ 
They are both minimal for the same value of the variational parameter
$\be=\be_1$ which solves the equation (\ref{quadraticequation}).
The eigenvalues of the hermitian matrix defined by (\ref{diagone}-\ref{nondiag}) 
are readily obtained :
\beqa
\tilde{E}_{\pm}=F+E_{cm}+\lan u_1|V_{eff}|u_1\ran \pm\nonumber\\ 
\sqrt{\fr{\B^2}{4}+|\lan u_1|V_{eff}|u_{-1}\ran|^2}.
\eeqa
The degeneracy of the states $m=\pm 1$ at zero field has been eliminated:
there is a gap of $2|\lan u_1|V_{eff}|u_{-1}\ran|$. Moreover,
the linear term $\pm \B/2$ is no longer present. Instead, there is a term
$\sqrt{\fr{\B^2}{4}+|\lan u_1|V_{eff}|u_{-1}\ran|^2}$ quadratic for small
$\B$. The lowest triplet energy $\tE_{-}=\tE_{\pm 1}$ is plotted in Fig. \ref{ddot}.
At zero field it is only about $0.4\:Ry$ higher than the numerical value.
As the densities of $m=\pm 1$ states are not peaked in the barrier, it is not
a surprise that the accuracy of our calculations is better than in the previous
case. The discrepancy increases, however, with the magnetic field. To improve upon our approximation
we take also $m=\pm 3$ states into account. The resulting energy $\tE_{\pm 1,\pm 3}$ does not 
differ much from the previous one at $\B=0$, but a remarkable accuracy was achieved
at higher fields (see Fig. \ref{ddot}). As before, we plot electronic densities at 
$\B=0\:Ry$ and $1\:Ry$. (See Fig. \ref{densitiesvt}, 
Fig. \ref{densitiesnt}). The low triplet density in the center
of the double dot is inherited from the single dot. Loosely speaking, the two electron
droplet in the lowest triplet state consists, from the outset, only of rotating components.  
After increasing the barrier and applying a magnetic field it will acquire components 
which rotate faster. As contrasted to the lowest singlet state, one can obtain a qualitatively 
valid description neglecting vibrations. This distinction between the lowest singlet and triplet     
states is likely to remain valid in more realistic double dots. It  may find applications
in spectroscopic measurements sensitive to charge distribution (e.g. in quantum point contact
measurements).

\subsection{Singlet-triplet transition}

\begin{figure}
\begin{center}
\includegraphics[angle=0,width=0.4\textwidth]{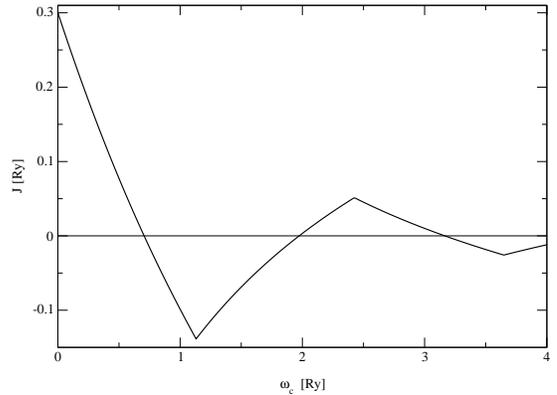}  
\caption{The singlet-triplet gap as a function of the magnetic field in a single dot with
$\om_0=\fr{2}{3}\:Ry$. Variational calculations.\\ }
\label{sdotJ}
\end{center}
\end{figure}

\begin{figure}
\begin{center}
\includegraphics[angle=0,width=0.4\textwidth]{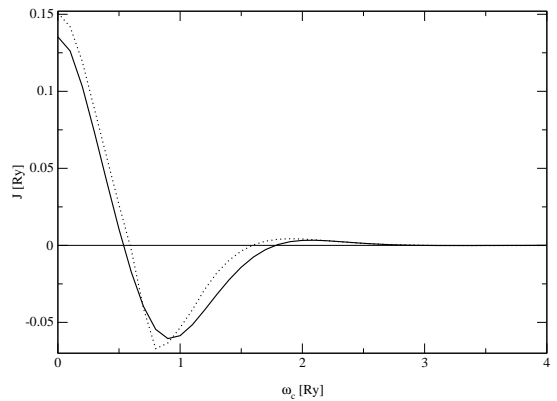}  
\caption{The singlet-triplet gap as a function of the magnetic field in a double quantum dot with
$\omega_0=\fr{2}{3}\:Ry$, $V_0=1\:Ry$, $\Delta=0.5\:a_B$. Solid line: variational
results. Dotted line: numerical calculations by the configuration-interaction
method.}
\label{ddotJ}
\end{center}
\end{figure}

We define the singlet-triplet gap $J:=E_t-E_s$, where $E_t$ ($E_s$) 
denotes the energy of the lowest lying triplet (singlet) and plot the $J(\B)$ dependence 
for a single dot ($\om_0=\fr{2}{3} \:Ry $, $V_0=0$) in Fig. \ref{sdotJ}.
Next, we move to the case of a double dot setting the parameters as in the previous 
subsections: $\om_0=\fr{2}{3}\:Ry$, $V_0=1\:Ry$, $\Delta=0.5 a_B$ and choose 
$E_t=\tE_{\pm 1,\pm 3}$, $E_s=\tE_{0,\pm 2}$. The respective $J(\B)$ function, 
plotted in Fig. \ref{ddotJ}, compares well with numerical results
which are also presented.

In Section II we argued that the increase of the magnetic field results in an increase of 
ground state angular momentum leading to singlet-triplet oscillations of two electrons
in a single dot. Although in a double dot the eigenstates do not have a definite angular 
momentum, the orbital Zeeman term and Coulomb interaction are still responsible for singlet-triplet 
transitions. In view of this fact and the discussion from the previous subsection on mixing of angular
momenta, it is not a surprise that the barrier only smoothed out the sharp edges of the
$J(\B)$ function which were a direct consequence of angular momentum conservation.   
In particular, the positions of the first crossing and minimum are similar as without the 
barrier.

Since the precise shape of the dot does not play a role in the 
above discussion it seems to us possible that these qualitative features of the $J(\B)$
dependence (the crossing, the minimum and the signature of the angular momentum
conservation) are independent of the confining potential. (See e.g. \cite{JMM}
on elliptical dots). This claim is also reinforced by a recent analysis
by Scarola and Das Sarma \cite{Scarola} who related the singlet-triplet
transitions to changes in vorticity of the two-electron wavefunction. Although 
their model potential and variational wavefunctions differ from ours, their
results reflect the features mentioned above. Consequently, we expect that also
realistic quantum dots of irregular shape will exhibit these properties.
On the other hand, more detailed features, like the value of $J(\B)$, the number of crossings or their precise positions will certainly depend on the shape of the dot.

\section{Conclusion}
In this paper we discussed the problem of the transition from an
artificial helium atom to a hydrogen molecule as a function
of the barrier potential. We illustrated it with a 
simple calculation of  the lowest lying singlet
and triplet states in the strong coupling limit, where the Heitler-London approach 
is not valid. To achieve this goal we introduced variational wavefunctions which 
describe accurately two Coulomb interacting electrons in a parabolic quantum dot. 
The singlet and triplet energies, and the singlet-triplet gap $J$, 
 were calculated
as a function of the barrier potential and the magnetic field. The origin of
the singlet-triplet transition was discussed.
We hope that these variational functions will 
also be useful in developing methods of isolating spins of individual 
electrons and coupling them in a controllable 
manner in lateral quantum dots, as well as 
in other  areas of research, e.g. to study with
analytical expressions the formation of 
Wigner molecules \cite{PSN,Uzi1}.  

We add two remark of a technical nature: First, in the case of a double quantum dot
we were using the parameters $\be$ optimized in a single circular dot. One could 
as well optimize them in the potential under study without much additional effort.   
It turns out, however, that the gain in accuracy is very small (at least in the case of
the double dot presented here) so we did not pursue this approach. Second, a
more realistic model of a double dot would be a single elliptical dot perturbed
by a barrier. Since an elliptical confinement still separates the center of mass 
and relative motion it is not difficult to adapt our method to this case.
Quantitative agreement with numerical results would require, however, more 
effort as our variational wavefunctions are optimized in a circular dot. On the
other hand, our approach works well in the case of a single elliptical dot 
(no barrier) or a quantum ring (a single circular dot with a circular barrier).

\begin{acknowledgments} 
The authors acknowledge support by the Institute
for Microstructural Sciences, NRC, by the 
National Science and Engineering Research Council, and by the Canadian Institute for Advanced Research. One of the authors (W.D.) also acknowledges support from the
EC Research Training Network 'Quantum Spaces - Non-commutative Geometry'.
The authors  thank M. Pioro-Ladriere, A. S. Sachrajda, M. Abolfath, C. Dharma-wardana, and W. Jakobiec for stimulating discussions during the course of this work.  
\end{acknowledgments}

\begin{appendix}

\section*{Appendix: The effect of radial excitations on the density in a double dot}

\begin{figure}
\begin{center}
\includegraphics[angle=0,width=0.4\textwidth]{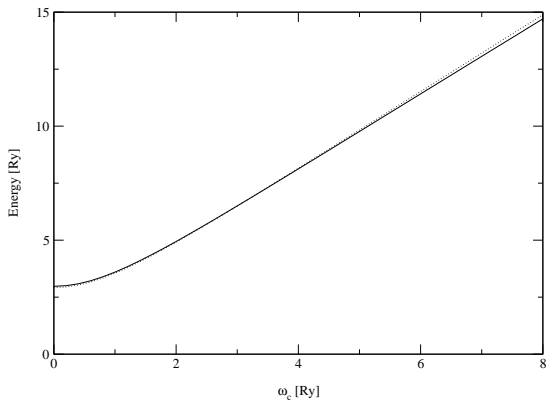}  
\caption{The energy of the first excited state in a single parabolic dot with
$\omega_0=\fr{2}{3}\:Ry$. Solid line: variational calculation. 
Dotted line: numerical calculations by the Numerov method.}
\label{excitedenergy}
\end{center}
\end{figure}

In this Appendix we calculate the two-electron density of the lowest singlet in a 
double dot taking into account the first excited state of the radial motion with $m=0$.
To this end, we proceed as follows: We start from a certain exact solution of the single dot 
problem. Then we construct a variational
wavefunction of a similar functional form and orthogonalize it to the ground state 
variational wavefunction. Finally, we diagonalize the double dot Hamiltonian using the
excited state $m=0^*$ together with the previously studied $m=0$, $m=2$, $m=-2$ states.

From the analysis by Taut \cite{Taut2} we obtain that at $\al=\fr{1}{12}$ there is
an eigenstate of energy $E=1\:Ry$ given by:
\beq
\tilde{u}_0(\rho)=\tilde{C}_0\rho^{\fr{1}{2}}(1+\sqrt{12}\rho+2\rho^2)e^{-\fr{\rho^2}{2}}.
\eeq 
As a matter of fact, at zero magnetic field it is the ground state, since it is positive. But  
the binding potential corresponding to this eigenstate is much weaker than the one we have
under study. We expect that at a stronger binding potential the first excited state with $m=0$ will
have such functional form so we describe it  by a variational wavefunction: 
\beq
\tu(\rho)=\tC\rho^{\fr{1}{2}}(1+\tbe\rho+\tde\rho^2)e^{-\fr{\rho^2}{2}},
\eeq 
where $\tC$ is a normalization constant, $\tbe$, $\tde$ are variational parameters.
First of all, we have to make sure that the state is orthogonal to $u_{0,\be_0}$:
\beqa
0=\int_0^{\infty}\tu(\rho)u_{0,\be_0}(\rho)d\rho=\nonumber\\
\fr{1}{2}(1+\fr{1}{2}\spi(\be_0+\tbe)+(\tde+\be_0\tbe)+\fr{3}{4}\spi\tde\be_0).
\eeqa
This implies that $\tde=-\Dj-\Dd\tbe$, where
\beqa
\Dj=\fr{1+\fr{1}{2}\spi\be_0}{1+\fr{3}{4}\spi\be_0},\\
\Dd=\fr{\fr{1}{2}\spi+\be_0}{1+\fr{3}{4}\spi\be_0}.
\eeqa
Next, we evaluate the normalization constant:
\beqa
\tC^{-2}&=&\tilde{d}+\tilde{e}\tbe+\tilde{f}\tbe^2,\\
\tilde{d}&=&\Dj^2-\Dj+\fr{1}{2},\\
\tilde{e}&=&\fr{1}{2}\spi-\fr{3}{4}\spi\Dj+2\Dj\Dd-\Dd,\\
\tilde{f}&=&\fr{1}{2}-\fr{3}{4}\spi\Dd+\Dd^2.
\eeqa
Now we are ready to calculate the variational energy:

\begin{widetext}
\beqa
\tilde{E}_0(\tbe)&=&2\al\int_0^\infty \tu(\rho)
\big\{-\pa_{\rho}^2+\big[\fr{1}{\sal\rho}+\rho^2-
\fr{1}{4\rho^2}\big]\big\}\tu(\rho)d\rho,\\
\tilde{E}_0(\tbe)&=&2\al\fr{\tilde{a}+\tilde{b}\tbe+\tilde{c}\tbe^2}
{\tilde{d}+\tilde{e}\tbe+\tilde{f}\tbe^2},\quad\label{aenergy}\\
\tilde{a}&=&(\fr{1}{2}\sqrt{\fr{\pi}{\al}}+1)-2\Dj(\fr{1}{4}\sqrt{\fr{\pi}{\al}}+1)
+\Dj^2(\fr{3}{8}\sqrt{\fr{\pi}{\al}}+4),\\
\tilde{b}&=&(\fr{1}{\sqrt{\al}}+\spi)-2\Dd(1+\fr{1}{4}\sqrt{\fr{\pi}{\al}})-
\Dj(\fr{5}{2}\spi+\fr{1}{\sqrt{\al}})+2\Dj\Dd(4+\fr{3}{8}\sqrt{\fr{\pi}{\al}}),\qquad\\
\tilde{c}&=&(\fr{1}{4}\sqrt{\fr{\pi}{\al}}+\fr{3}{2})-\Dd(\fr{5}{2}\spi+\fr{1}{\sqrt{\al}})
+\Dd^2(4+\fr{3}{8}\sqrt{\fr{\pi}{\al}}).
\eeqa
\end{widetext}
The optimal variational parameter $\tbe_0$ can be found by solving a quadratic equation
analogous to (\ref{quadraticequation}). The energy is then obtained substituting it back
to equation (\ref{aenergy}). At this stage a comparison can be made between our
variational approach and a numerical solution of the single dot problem by the Numerov method.
The results are plotted in Fig. \ref{excitedenergy}. Similarly as in Section II, we 
note very good agreement between the two methods. Finally, we arrive at a task of 
computing matrix elements of the effective potential $V_{eff}$:  

\begin{figure}
\begin{center}
\includegraphics[angle=-90,width=0.4\textwidth]{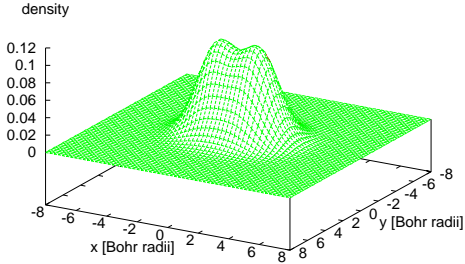}  
\caption{The two-electron density at $\B=0\:Ry$ calculated using $m=0$, $m=0^*$, $m=2$, $m=-2$ states.}
\label{exciteddensity}
\end{center}
\end{figure}

\begin{widetext}
\beqa
\lan \tuo|V_{eff}|\tuo\ran &=& \fr{V_0\ga\tCo^2}{\sqrt{\ga^2+1}}
\big(K(\ga^2+1,1,0)+(\tbe_0^2+2\tde)K(\ga^2+1,2,0)+ \nonumber \\ & &
\tde^2K(\ga^2+1,3,0)+2\tbe_0K(\ga^2+1,\fr{3}{2},0)
+2\tbe\tde K(\ga^2+1,\fr{5}{2},0)\big),\\
\lan\tuo|V_{eff}|u_{m}\ran &=&
\fr{V_0\ga\tCo C_{m,\be_m}}{\sqrt{\ga^2+1}}\big(K(\ga^2+1,\fr{2+|m|}{2},m) 
+(\be_m+\tbe_0)K(\ga^2+1,\fr{3+|m|}{2},m)+\nonumber\\ & &(\tbe_0\be_m+\tde)K(\ga^2+1,\fr{4+|m|}{2},m)
+\tde\be_m K(\ga^2+1,\fr{5+|m|}{2},m)\big).\quad
\eeqa
\end{widetext}

(The function $K$ was defined in Section III.) 
Together with matrix elements from Section III we have all input necessary to 
diagonalize the double dot Hamiltonian in the subspace spanned by the 
$m=0$, $m=0^*$, $m=2$, $m=-2$ states (where $0^*$ denotes the first excited state
with $m=0$, determined in this Appendix). Having obtained the wavefunction, we calculate
the density at $\B=0$ and plot it  in Fig. \ref{exciteddensity}. 
Comparison with the density at $\B=0$ calculated previously
(Fig. \ref{densitiesv}) indicates that the radial motion excitation lowers the density 
in the center of the double dot.

\end{appendix}


\end{document}